\documentclass[reprint, letterpaper, nobibnotes, aps, superscriptaddress,prb]{revtex4-1}

\usepackage{etoolbox}
\usepackage{graphicx}
\usepackage{amssymb}
\usepackage{amsmath}
\usepackage{color}
\usepackage{url}
\usepackage{multirow}
\usepackage{graphicx}        

\newcommand{\arxiv}{\let\arxivFig\matt}

\def\Xint#1{\mathchoice
   {\XXint\displaystyle\textstyle{#1}}%
   {\XXint\textstyle\scriptstyle{#1}}%
   {\XXint\scriptstyle\scriptscriptstyle{#1}}%
   {\XXint\scriptscriptstyle\scriptscriptstyle{#1}}%
   \!\int}
\def\XXint#1#2#3{{\setbox0=\hbox{$#1{#2#3}{\int}$}
     \vcenter{\hbox{$#2#3$}}\kern-.5\wd0}}

\def\dashint{\Xint-}

\newtoggle{fulltitlepage}
\settoggle{fulltitlepage}{true}

\begin{document}

\title{Modeling Charge-Sign Asymmetric Solvation Free Energies With Nonlinear Boundary Conditions}

\iftoggle{fulltitlepage}{
\author{Jaydeep P. Bardhan}
\affiliation{Dept.~of Electrical and Computer Enginering, Northeastern University, Boston MA 02115}
\author{Matthew G. Knepley}
\affiliation{Computation Institute, The University of Chicago, Chicago IL 60637}
\begin{abstract}
We show that charge-sign-dependent asymmetric hydration can be modeled
accurately using linear Poisson theory but replacing the standard
electric-displacement boundary condition with a simple nonlinear
boundary condition.  Using a single multiplicative scaling factor to
determine atomic radii from molecular dynamics Lennard-Jones
parameters, the new model accurately reproduces MD free-energy
calculations of hydration asymmetries for (i) monatomic ions, (ii)
titratable amino acids in both their protonated and unprotonated
states, and (iii) the Mobley ``bracelet'' and ``rod'' test problems
[J. Phys. Chem. B, v. 112:2408, 2008].  Remarkably, the model also
justifies the use of linear response expressions for charging free
energies.  Our boundary-element method implementation demonstrates the
ease with which other continuum-electrostatic solvers can be extended
to include asymmetry.
\end{abstract}
}{
}

\maketitle

\section{Introduction}\label{sec:intro}
Implicit-solvent models represent an intuitive and fast approach to
understand molecular solvation~\cite{Roux99,Sharp90,Davis90,Tomasi94}, and
have a rigorous statistical-mechanical interpretation as an
approximation to the potential of mean force (PMF) experienced by a
molecular solute due to the surrounding solvent
molecules~\cite{Roux99}.  The PMF is usually decomposed into non-polar
and electrostatic terms, the latter of which are often modeled using
macroscopic continuum models based on the Poisson--Boltzmann
partial-differential equation (PDE).  Continuum models approximate the
free energy required to grow the solute charge distribution into the
solute
cavity~\cite{Roux99,Kirkwood34,Latimer39,Kornyshev97,Hildebrandt04,Rizzo06,Abrashkin07,Gong09,Guo13,Zhou14}.
Although implicit-solvent models can be orders of magnitude faster
than explicit-solvent molecular-dynamics (MD) simulations, most
popular continuum theories ignore numerous potentially important
effects, including solvent molecules' finite size and specific
molecular interactions such as hydrogen bonding (the AGBNP2 model,
which addresses the latter, is a notable
exception~\cite{Gallicchio09}).

One of the Poisson model's most perplexing and long-standing
shortcomings is the difficulty of extending it to model charge-sign
asymmetric solvation: for example, given two monatomic ions of equal
radius, one of $+q$ charge and the other of $-q$, the negative charge
experiences stronger interactions with the solvent (more negative
solvation free
energy)~\cite{Latimer39,Rashin85,Ashbaugh00,LyndenBell01,Rajamani04,Grossfield05,Mukhopadhyay12,Bardhan12_asymmetry}.
However, standard Poisson models are charge-sign \textit{symmetric};
that is, they predict the same solvation free energy for $\pm q$.  The
need to include asymmetric effects is difficult to exaggerate,
particularly in biological contexts.  Consider that the protein avidin
binds its ligand biotin with a binding free energy of approximately
$-$20~kcal/mol, one of the most favorable in biology~\cite{Green75};
solvent-exposed $+1e$ and $-1e$ charges can experience as much as
40~kcal/mol difference in their solvation free
energies~\cite{Bardhan12_asymmetry}.  Dominant factors in charge-sign
asymmetric response include the liquid-vapor interface
potential~\cite{Ashbaugh00,Rajamani04} and the fact that water
hydrogens can approach a negative solute charge closer than water
oxygens can approach a positive
one~\cite{Latimer39,Rashin85,Mukhopadhyay12,Bardhan12_asymmetry}.
Spherical solutes with central charges provide a useful data set for
developing an understanding of size- and charge-sign dependent
hydration, including the characterization of interface potentials,
solvent packing, and dielectric
saturation~\cite{Rashin85,Hummer96_netchargecorrection,Ashbaugh00,Rajamani04,Grossfield05,Fedorov07}.
These analyses and the continuum macroscopic-dielectric framework
suggest that improvements require a more detailed, accurate
representation of the solvent dipole field
$\mathbf{P}(\mathbf{r})$~\cite{Warshel76,Azuara08}, or, equivalently,
the solvent charge density ~\mbox{$\rho_{\mathrm{induced}}(\mathbf{r})
  = \nabla \cdot \mathbf{P}(\mathbf{r})$}.  Because $\mathbf{P}$ and
$\rho_{\mathrm{induced}}$ do \textit{not} respond linearly to the
solute charge distribution~\cite{Alper90}, particularly in the first
solvent shell~\cite{Purisima09}, many groups have developed solvent
models in which the solvent potential obeys a nonlinear partial
differential equation
(PDE)~\cite{Kornyshev98,Sandberg02_first_LD,Jha08,Gong08,Hu12_Wei_nonlinear_Poisson_BJ}.
Unfortunately, most of these models are still charge-sign symmetric.

However, in 1939 Latimer et al. proposed an approach to increase or
decrease an ion's radius based on the charge~\cite{Latimer39}, and
recent developments in high-performance computing and explicit-solvent
MD free-energy calculations provide important new data to extend this
approach.  Mobley et al. constructed a challenging test set and
conducted extensive MD simulations on charge-sign
asymmetry~\cite{Mobley08_asymmetry}, enabling important new
developments in modeling
asymmetry~\cite{Purisima09,Corbeil10,Mukhopadhyay14} that extend
Latimer's work to Generalized-Born (GB) models of complex solutes.  GB
theory was a natural setting for these developments because Latimer's
work and GB theory share the conceptual picture of an effective atomic
radius.  These early studies provided an important insight: a buried
charge still affects the electric field at the boundary, so merely
parameterizing charge-dependent radii cannot (indeed, should not)
provide a satisfactory explanation.  The accuracy of asymmetric GB
models suggests that a simple Poisson-based model exists, but finding
one has proven to be surprisingly difficult.


In this paper, we propose a simple Poisson continuum model that
includes charge-sign asymmetry and show that it is remarkably accurate
even without parameterization on an atom-by-atom basis.  The key
feature of our theory is a \textit{nonlinear boundary condition}
(NLBC) for the normal displacement field; in contrast, the
displacement boundary condition for the standard (symmetric) Poisson
theory is linear~\cite{Jackson_classical_electrodynamics}.
Importantly, even though our proposed displacement boundary condition
is nonlinear, the electrostatic potential in the solvent and solute
volumes still satisfy \textit{linear} Poisson/Laplace equations.  Two
phenomena motivated us to propose a nonlinear boundary condition
instead of a nonlinear governing equation.  First, numerous results
illustrate that the solute reaction potential obeys nearly linear
response even though the solvent charge distribution does
not~\cite{Lin11_Pettitt,Boda09,Purisima09,Corbeil10,Mukhopadhyay14,Bardhan12_asymmetry}.
For example, the new asymmetric Generalized-Born (GB) models use the
charge distribution only to modify the Born radii, with the overall
energy still computed using superposition (independent sum of
individual charge
responses)~\cite{Purisima09,Corbeil10,Mukhopadhyay14}.  Furthermore,
we found in our previous work that the solute reaction potential is
essentially a \textit{piecewise-linear} function of charge
~\cite{Bardhan12_asymmetry}, i.e. the proportionality coefficient
depends on whether one is charging an ion from zero to $+q$ or to
$-q$.  In fact, we began this work seeking primarily to reproduce this
curiously simple nonlinearity.

The second phenomenon motivating our NLBC approach is the fact that
the solute reaction potential is a harmonic field---that is, it
satisfies the Laplace equation.  This property is useful for numerical
computations~\cite{Chern03,Holst12} and also provides a path to
improve models via boundary-integral methods~\cite{Bardhan12_review}:
harmonicity means that regardless of the solvent model of interest,
there exists \textit{some} surface charge density that reproduces the
reaction potential inside.  For a given solvent model, the surface
charge density might satisfy a nonlinear boundary-integral equation,
but the very fact that such a density always exists suggests that one
might improve continuum models by adding nonlinear terms to widely
used BIE formalisms~\cite{Rizzo67,Shaw85,Juffer91,Bardhan09_disc}. 


\section{Continuum Model and Extension to Nonlinear Boundary Conditions}
We first present the standard (charge-sign symmetric) Poisson
electrostatic model and then describe the difference between it and
our proposed NLBC model.  In both theories, the molecular solute is
treated as a macroscopic linear dielectric continuum obeying the
Poisson equation \mbox{$\nabla^2 \varphi_1(\mathbf{r}) =
  -\frac{\rho(\mathbf{r})}{\epsilon_1}$} where $\mathbf{r}$ is a point
in space, $\varphi_1(\mathbf{r})$ is the potential in the solute,
$\epsilon_1$ is the relative permittivity, and the molecular charge
distribution $\rho(\mathbf{r})$ is a set of $N_q$ point charges, i.e.
\mbox{$\rho(\mathbf{r}) = \sum_{i=1}^{N_q} q_i
  \delta(\mathbf{r}-\mathbf{r}_i)$}.  The solute and solvent are
separated by the interface $\Gamma$, and the solvent exterior is a
linear dielectric with permittivity \mbox{$\epsilon_2 \gg
  \epsilon_1$}, so the electric potential obeys \mbox{$\nabla^2
  \varphi_2(\mathbf{r}) = 0$}; note that modeling realistic biological
solutions requires inclusion of screening effects due to mobile ions
using e.g. some form of the Poisson--Boltzmann equation for
$\varphi_2(\mathbf{r})$~\cite{Sharp90,Tomasi94}.  From macroscopic
dielectric theory and Gauss's law, we obtain the standard Maxwell
boundary conditions for $\mathbf{r}_\Gamma \in \Gamma$
\begin{align}
  \varphi_1(\mathbf{r}_\Gamma) &= \varphi_2(\mathbf{r}_\Gamma),\\
  \epsilon_1 \frac{\partial \varphi_1}{\partial n}(\mathbf{r}_\Gamma) &= \epsilon_2 \frac{\partial \varphi_2}{\partial n}(\mathbf{r}_\Gamma),\label{eq:standard-Maxwell-bc}
\end{align}
where $\frac{\partial}{\partial n}$ denotes the normal derivative (the
normal at $\mathbf{r}_\Gamma$ is defined pointing outward into
solvent). Assuming that $\varphi_2(\mathbf{r})$ decays sufficiently
quickly as $|\mathbf{r}|\rightarrow\infty$, this mixed-dielectric
Poisson problem is well posed and the unknown potential $\varphi_1$
can be rewritten as a linear boundary-integral equation for an unknown
surface charge distribution on $\Gamma$.  In particular, the
apparent-surface charge (ASC)
model~\cite{Shaw85,Altman05_2,Bardhan09_disc} (also known as the
polarizable continuum model~\cite{Miertus81,Tomasi94}) can be
interpreted as finding an equivalent surface charge
$\sigma(\mathbf{r})$ in a homogeneous medium with permittivity
$\epsilon_1$ everywhere.  In this equivalent problem, the
analogous boundary condition to Eq.~\ref{eq:standard-Maxwell-bc}
is simpler due to homogeneity, but adds a term for the surface charge:
\begin{equation}
\frac{\sigma(\mathbf{r}_\Gamma)}{\epsilon_1} = \frac{\partial
    \hat{\varphi}_1}{\partial n}(\mathbf{r}_\Gamma) - \frac{\partial
    \hat{\varphi}_2}{\partial n}(\mathbf{r}_\Gamma),\label{eq:ecf-bc}
\end{equation}
and we use $\hat{\varphi}_i = \varphi_i$ to emphasize our use of an
equivalent problem.  Defining \mbox{$G(\mathbf{r};\mathbf{r}') =
  \frac{1}{4 \pi || \mathbf{r} - \mathbf{r}'||}$}, one obtains
\begin{align}
  \left(I  + \hat{\epsilon} \left(-\frac{1}{2} I+ K\right)\right)\sigma &= -\hat{\epsilon}\sum_i^{N_q} q_i \frac{\partial G}{\partial n}
\end{align}
where $\hat{\epsilon} = (\epsilon_2-\epsilon_1)/\epsilon_2$ and $K$ is
the normal electric field operator~\cite{Bardhan09_disc}.  The
reaction potential in the solute is then
\mbox{$\varphi^{REAC}(\mathbf{r}) = \frac{1}{\epsilon_1} \int_\Gamma
  G(\mathbf{r}; \mathbf{r}') \sigma(\mathbf{r}') dA'$}, and
\mbox{$\varphi_{1}(\mathbf{r}) = \varphi^{REAC}(\mathbf{r})
  +\varphi^{Coulomb}(\mathbf{r})$}, with the latter term representing
the Coulomb potential due to $\rho(\mathbf{r})$.

The standard Maxwell displacement boundary condition
Eq.~\ref{eq:standard-Maxwell-bc} is obtained using Gauss's law in
integral form and the fact that the divergence of the polarization
field $\mathbf{P}(\mathbf{r})$ represents a volume charge density.
However, near the solute--solvent boundary, the assumption that
$\mathbf{P}(\mathbf{r})$ is pointwise proportional to the local
electric field breaks down due to water structure at the interface;
that is, it is no longer necessarily true that
\mbox{$\mathbf{P}(\mathbf{r}) = (\epsilon(\mathbf{r})-1)
  \mathbf{E}(\mathbf{r})$}.

To model nonlinear solvent response at the boundary, we propose to
replace the linear boundary condition,
Eq.~\ref{eq:standard-Maxwell-bc}, with the phenomenological nonlinear
boundary condition
\begin{equation}
f(E_n) \frac{\partial \varphi_1}{\partial n}(\mathbf{r}_\Gamma) = \left(1+f(E_n)\right)\frac{\partial \varphi_2}{\partial n}(\mathbf{r}_\Gamma)\label{eq:charge-layer-nonlinear-BC}
  \end{equation}
where $E_n$ is the electric field just inside $\Gamma$,
i.e. \mbox{$E_n = -\sum_i q_i \frac{\partial G}{\partial n} - K
  \sigma$}, and
\begin{align}
f(E_n) & = \frac{\epsilon_1}{\epsilon_2-\epsilon_1} - h(E_n);\\
h(E_n) & = \alpha \tanh(\beta E_n - \gamma) + \mu.\label{eq:tanh}
  \end{align}
with $\alpha$, $\beta$, and $\gamma$ representing model parameters and
\mbox{$\mu = -\alpha
  \tanh(-\gamma)$}. The specification of $\mu$ ensures that $h(E_n=0)
= 0$, so that in the limit of weak electric fields, such as induced at
the surface by a deeply buried charge, the boundary condition reduces
to the familiar Poisson model.  The NLBC leads to
the modified, \textit{nonlinear} BIE
\begin{equation}
\left(I + \hat{\epsilon}\left(-\frac{1}{2}I + K\right) + h(E_n)\right) \sigma = -\hat{\epsilon}\sum_{i} q_i \frac{\partial G}{\partial n},
  \end{equation}
with the nonlinearity arising in the dependence of $h$ on $E_n$ (see
the Supporting Information for details on the numerical
implementation).

One challenge in developing more accurate solvent models is the fact
that nonlinear response~\cite{Sharp90_2} generally requires a charging
process~\cite{Zhou94}, i.e. the expression \mbox{$\Delta G^{solv,es} = \frac{1}{2}
  q^T \varphi^{REAC} = \frac{1}{2}q^T L q$} no longer holds ($L$
denotes the reaction-potential
operator~\cite{Roux99,Bardhan12_review}). However, our previous work
showed the remarkable fact that the solute reaction potential is
\textit{piecewise} linear, with the breakpoint at
$q=0$~\cite{Bardhan12_asymmetry}, so that $\varphi^{REAC} = L_+ q$ for
$q > 0$ and $\varphi^{REAC} = L_- q$ for $q < 0$, with $L_+ \neq L_-$.
The proposed NLBC in Eqs.~\ref{eq:charge-layer-nonlinear-BC}
and~\ref{eq:tanh} immediately explains this curious phenomenon:
consider the limit $\beta \rightarrow \infty$, so that $\tanh$ is
constant everywhere, but discontinuous at $q=0$.  The Debye charging
process~\cite{Kirkwood34_2} scales all charges uniformly, i.e.
\mbox{$\hat{\rho}(r; \lambda) = \lambda \rho(r)$}, so the Coulomb
field \mbox{$\frac{\partial \varphi^{Coul}}{\partial n}$} has the same
sign for all finite $\lambda$. The Coulomb-field approximation (CFA)
shows that the reaction field is nearly proportional to the direct
Coulomb field, but slightly smaller in
magnitude~\cite{Kharkats76,Bardhan08_BIBEE,Bardhan12_review}, so for
finite $\lambda$, at almost all $\mathbf{r}_\Gamma$, the total field
$E_n(\mathbf{r}_\Gamma; \lambda)$ has the same sign as
$E_n(\mathbf{r}_\Gamma; \lambda = 1)$.  This implies that almost
everywhere on the surface, the $\tanh$ boundary condition takes its
limiting ($\lambda \rightarrow 1$) value for any finite $\lambda$,
which means that the boundary condition is essentially linear:
\mbox{$(1 + g(r))\sigma(r) = \frac{\partial \hat\varphi_2}{\partial
    n}-\frac{\partial \hat\varphi_1}{\partial n}$}.  With this
justification, in this work we compute solvation free energies as
\mbox{$\Delta G^{solv,es}= \frac{1}{2} q^T \varphi^{REAC}$}.  Note
that a more precise definition of the charging free energy would be
piecewise \textit{affine}, because the charging free energy also
includes a linear term that results from the liquid-vapor interface
potential~\cite{Harder08,Kathmann11,Bardhan12_asymmetry}; as noted
above, however, in the present work its influence is approximated via
the offset parameter $\gamma$.

\section{Results and Discussion}\label{sec:results}

We parameterized the NLBC model using the Mobley et al. MD free-energy
calculations, who studied asymmetry using fictitious bracelet and rod
molecules~\cite{Mobley08_asymmetry} constructed from AMBER
$\mathrm{C}_{\alpha}$ atoms with $\mathrm{R}_{\mathrm{min}}/2 = 1.908$~\AA.  We
obtained optimal results with $\alpha = 0.5$, $\beta = -60$, $\gamma =
-0.5$, and a continuum-model $\mathrm{C}_\alpha$ radius of 1.75~\AA~(a
scale factor of approximately 0.92). Note that in this first
exploration of the NLBC, we have parameterized against the overall
solvation free energies computed by Mobley et al. rather than the more
correct charging free energy.

\begin{figure}[ht!]
  \centering \resizebox{3.0in}{!}{\includegraphics{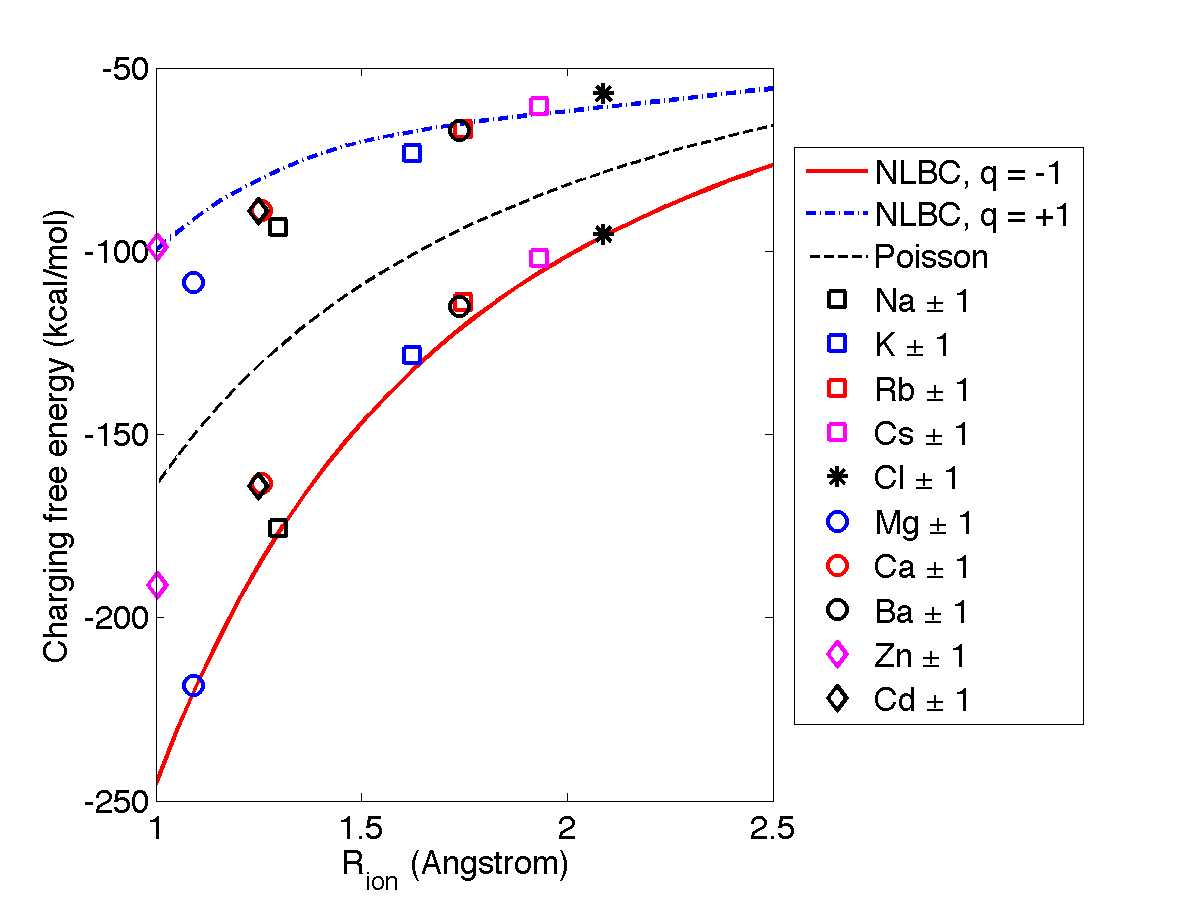}}
  \caption{Asymmetric polarization free energies for a monovalent
    central charge in a sphere, as a function of sphere radius.  The
    labeled symbols denote results from MD free-energy calculations
    charging CHARMM monatomic ions from zero to $+1e$ or $-1e$, with
    $\mathrm{R}_{\mathrm{ion}} = 0.92 \mathrm{R}_{\mathrm{min}}/2$.  The dashed black curve
    in the middle is the (charge-sign symmetric) Born polarization
    free energy.}\protect\label{fig:standard-ions-model} \end{figure}

Figure~\ref{fig:standard-ions-model} plots NLBC and MD free-energy
calculations for ion charging free energies; the MD charging
simulations used in our previous work~\cite{Bardhan12_asymmetry} (see
Supporting Information) and CHARMM Lennard-Jones parameters.  We
remind the reader that no additional parameters were fit in obtaining
these NLBC results, i.e. ion radii were assigned
$\mathrm{R}_{\mathrm{ion}} = 0.92 \mathrm{R}_{\mathrm{min}}/2$.  For
additional data, ions were charged to both $+1e$ and $-1e$, regardless
of the charge on the real ion, and the NLBC accurately predicts these
charging free energies as well.  The largest deviations occur for
radii less than 1.4~\AA, where discrete packing effects and actual
dielectric saturation are likely.

\begin{table}
  \centering
\begin{tabular}{l|cc|cc}
  Problem & \multicolumn{2}{c}{Solvation errors} &  \multicolumn{2}{c}{Asymmetry errors}\\\hline
  & RMSE &  Max.  & RMS & Max. \\
  Rods & 5.57 & 9.63 & 0.88 & 1.49  \\
  Bracelets (opposing) & 2.88 & 6.10 & 2.04 & 3.08  \\
  Bracelets (distributed) & 2.20 & 2.72 & 0.29 & 0.59 \\
  Bracelets (dipole) & 2.67 & 3.52 & 0.85 & 1.09 
\end{tabular}
\caption{Comparison of NLBC model to MD free-energy calculations of
  Mobley et al.~\cite{Mobley08_asymmetry} for rod and bracelet
  molecule test set.  All energies are in kcal/mol.  See Supporting
  Information for detailed results.}\protect\label{table:Mobley}
\end{table}
Calculations for the Mobley test set are summarized in
Table~\ref{table:Mobley}; SI Figures~1--8 plot the NLBC and Mobley MD
solvation free energies and asymmetry energies, and include
illustrations of the test problems.  The rod molecules are composed of
5 or 6 atoms along a line, with one atom possessing $+1e$ charge, one
with $-1e$, and the rest neutral.  The asymmetry errors in
Table~\ref{table:Mobley} represent the difference in solvation
energies when reversing the charged atoms' signs.  The bracelet
molecules are regular polygons with between 3 and 8 sides; atoms are
at the vertices (1.4~\AA~apart). Bracelets were simulated with three
charge distributions: the ``opposing'' case had a $+1e$ charge
neutralized by two $-0.5e$ charges positioned symmetrically on the
opposite side.  The ``distributed'' case has one $+1e$ charge and a
neutralizing $-1e$ distributed equally on all the other atoms; the
``dipole'' case is similar to ``opposing'' but fixes the dipole
moment~\cite{Mobley08_asymmetry}.  Solvent charge-densities from the
MD calculations~\cite{Mobley08_asymmetry} suggest that solvent packing
may be responsible for size-dependent deviations; parameterizing radii
for actual atoms should significantly reduce these errors.

To test the model on real but nonspherical molecules, we compared NLBC
and MD charging free energies for isolated titratable amino acids in
both protonated and unprotonated states (See Supporting Information
for details on structure preparation). Parameters were from the CHARMM
force field~\cite{Brooks83} when available, with other protonation
states defined so that the protonated and unprotonated states had the
same number of atoms. The MD free-energy-perturbation (FEP)
calculations used the same protocol as the
ions~\cite{Bardhan12_asymmetry}, holding the solute rigid so that
$\epsilon_1 = 1$ unambiguously~\cite{Roux99}.  The deviations between
our MD results and the MD calculations of Nina et al.~\cite{Nina97}
are small compared to the energies of interest, and likely due to our
use of (i) periodic boundary conditions, (ii) a larger solvent box
(1959 waters vs. 150), and (iii) slightly different backbone angles.

\begin{figure}[ht!]
  \centering \resizebox{3.0in}{!}{\includegraphics{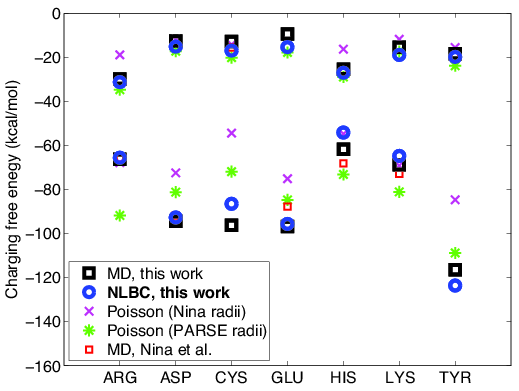}}
  \caption{Comparison of NLBC model to explicit-solvent MD FEP
    calculations for titratable residues with neutral blocking
    groups. MD results from Nina et al.~\cite{Nina97} are shown where
    available; standard continuum model results are shown for Nina et
    al. radii and PARSE radii~\cite{Sitkoff94} (using CHARMM
    charges).}\protect\label{fig:residues}
\end{figure}

As in the ion and Mobley examples, the NLBC radii were defined by the
scaling $\mathrm{R} = 0.92 \mathrm{R}_{\mathrm{min}}/2$. The results
in Figure~\ref{fig:residues} illustrate that the NLBC model correctly
captures solvation free energies in both charge states, despite the
fact that radii were not adjusted individually or even for the atomic
charges.  In contrast, standard Poisson model results computed using
the Nina et al.~\cite{Nina97} or PARSE~\cite{Sitkoff94} radii exhibit
larger deviations, particularly for arginine, aspartic acid, cysteine,
glutamic acid, and tyrosine.  These data suggest that the differences
between symmetric and asymmetric electrostatic models are robust with
respect to radii (the PARSE calculations are merely suggestive because
these calculations used the CHARMM charges; for consistent comparison
to experiment, one should use PARSE charges with PARSE radii).

\section{Conclusion}\label{sec:conclusion}

We have proposed a Poisson-based theory that models
charge-sign-dependent asymmetries in electrostatic solvation free
energies using a nonlinear boundary condition (NLBC), while still
using linear continuum theory in the solute and solvent volumes.  The
NLBC model accurately reproduces MD free-energy results for monatomic
ions, the Mobley et al. bracelet and rod problems, and titratable
residues, even though we have used charge-independent radii that were
fixed by a single scaling factor applied to MD radii.  Furthermore,
the NLBC reduces smoothly to the standard Poisson model as the
parameter $\alpha$ approaches zero.  Finally, our boundary-element
method implementation for non-trivial molecules demonstrates that the
new model is easily implemented in numerical Poisson and
Poisson--Boltzmann solvers.

Our introduction of a modified boundary condition to account for
solvation-shell response follows a long history in continuum
mechanics, where phenomenological techniques find applications in many
areas of science and engineering to capture a particular physical
behavior in continuum theory rather than modeling or deriving it from
first
principles~\cite{MizziBarberEmersonReeseStefanov2007,Brenner2011,Bardhan11_Knepley}. Non-equilibrium
micro-scale gas flows offer a well-developed example: velocity-slip
and temperature-jump boundary conditions are simplified
phenomenological approaches to represent both non-equilibrium and
gas-surface interaction effects occurring near solid walls. Such
boundary conditions were first suggested in the 19th century by
Maxwell~\cite{Maxwell1878} and von
Smoluchowski~\cite{Smoluchowski1898}, respectively.  More recent
examples include the partitioning of minerals at phase boundaries in
geophysics~\cite{LHeureux96}, tumor growth~\cite{Macklin05}, the
deformation of biological membranes~\cite{Fan03}, and thin electric
double layers in electro-osmotic flow~\cite{Yossifon07}.

Much as Beglov and Roux showed that solvent response approaches the
linear Poisson model in the limit as the solvent molecule approaches
zero size~\cite{Beglov96}, our model emphasizes that the nonlinear
response is generally localized in the first solvent shell.
Conceptually, the nonlinear boundary condition penalizes negative
surface charge because the larger water oxygen cannot approach a
solute charge as closely as the water hydrogens can.  From a
boundary-integral point of view, this has the same effect as adjusting
the atomic radii, an approach pioneered by Latimer et
al.~\cite{Latimer39}, and extended recently to GB
models~\cite{Purisima09,Corbeil10,Mukhopadhyay12,Mukhopadhyay14}.
Purisima's work is particularly relevant due to their use of
surface-charge boundary-integral approach, adjusting GB radii using
$\sigma(\mathbf{r})$~\cite{Purisima09,Corbeil10}.  Our work differs
substantially from these approaches because we have included asymmetry
directly in the underlying Poisson model.

The present theory can be extended in several important ways.  First,
the proposed NLBC model has only three parameters whose particular
dependencies on solvent model have not yet been established
theoretically.  Second, it seems straightforward to include ionic
screening via the Poisson--Boltzmann equation.  Third, the proposed
NLBC depends exclusively on the normal electric field; improved models
might include local curvature or higher-order moments of the
potential.  Importantly, the latter could distinguish between
small-magnitude charges near the surface, and larger charges further
away~\cite{Mukhopadhyay12}.  Fourth, water's length-scale-dependent
dielectric behavior might be included using nonlocal
electrostatics~\cite{Hildebrandt04,Fedorov07,Bardhan11_pka,Bardhan11_DAC,Bardhan13_nonlocal_review}. The
new model also does not necessarily capture specific hydrogen-bonding
effects like AGBNP2 does~\cite{Gallicchio09}, which motivates future
work comparing the two approaches.


\iftoggle{fulltitlepage}
{
\section*{Acknowledgments}
The authors thank David Mobley for sharing detailed calculation
results, Jed Brown and David Green for valuable discussions, and Matt
Reuter for a critical reading of the manuscript. MGK was partially
supported by the U.S. Department of Energy, Office of Science,
Advanced Scientific Computing Research, under Contract
DE-AC02-06CH11357, and also NSF Grant OCI-1147680. JPB has been
supported in part by the National Institute of General Medical
Sciences (NIGMS) of the National Institutes of Health (NIH) under
award number R21GM102642. The content is solely the responsibility of
the authors, and does not necessarily represent the official views of
the National Institutes of Health.

\section*{Supporting Information}
Figures comparing NLBC and MD calculations for the Mobley test
set~\cite{Mobley08_asymmetry}.  The source code (MATLAB) and surface
discretizations for running the nonlinear boundary-condition
calculations, data files, parameters, and scripts for preparing and
running the MD calculations of titratable residues, as well as source
code to generate the figures, are freely and publicly available online
at \url{https://bitbucket.org/jbardhan/si-nlbc}.
}{
}

\bibliographystyle{unsrt}
\bibliography{implicit-review}

\end{document}